\documentclass[prd,aps,showpacs,preprintnumbers,amsmath,amssymb,
nofootinbib,floatfix]{revtex4}
\usepackage{graphicx}
\usepackage{bm}

\newcommand{\be}{\begin{equation}}
\newcommand{\ee}{\end{equation}}
\newcommand{\bea}{\begin{eqnarray}}
\newcommand{\eea}{\end{eqnarray}}

\begin{document}

\title{Searching for Signatures of Cosmic Superstrings in the CMB} 

\date{\today}

\author{Rebecca J. Danos}\email[email: ]{rjdanos@hep.physics.mcgill.ca}

\affiliation{Department of Physics, McGill University, 
Montr\'eal, QC, H3A 2T8, Canada} 

\author{Robert H. Brandenberger}\email[email: ]{rhb@physics.mcgill.ca}

\affiliation{Department of Physics, McGill University, 
Montr\'eal, QC, H3A 2T8, Canada} 

\begin{abstract}

Because cosmic superstrings generically form junctions and gauge theoretic
strings typically do not, junctions may provide a signature to distinguish
between cosmic superstrings and gauge theoretic cosmic strings. In
cosmic microwave background anisotropy maps, cosmic strings lead to
distinctive line discontinuities. String junctions lead to junctions
in these line discontinuities. In turn, edge detection algorithms such
as the Canny algorithm can be used to search for signatures of strings
in anisotropy maps. We apply the Canny algorithm to simulated maps
which contain the effects of cosmic strings with and without string junctions.
The Canny algorithm produces edge maps. To distinguish between
edge maps from string simulations with and without junctions, we
examine the density distribution of edges and pixels crossed by edges.
We find that in string simulations without Gaussian noise (such as
produced by the dominant inflationary fluctuations) our analysis
of the output data from the Canny algorithm can clearly distinguish
between simulations with and without string junctions. In the presence
of Gaussian noise at the level expected from the current bounds on the
contribution of cosmic strings to the total power spectrum of density
fluctuations, the distinction between models with and without junctions
is more difficult. However, by carefully analyzing the data the
models can still be differentiated. 

\end{abstract}

\pacs{}

\maketitle

\section{Introduction}

In recent years there has been a revival in interest in finding signatures
of cosmic strings. One of the main reasons is that a large
class of string inflationary models predict 
\cite{Sarangi:2002yt} a copious
production of cosmic superstrings 
\cite{Witten:1985fp} - cosmic strings consisting of 
extended fundamental strings of cosmological scale - 
at the end of the period of inflation. 
Under certain conditions \cite{Copeland:2003bj}, these
cosmic superstrings are stable on cosmological time scales. 

Since gauge theories also predict the presence of cosmic strings
\cite{Kibble} (see \cite{VilShell,HK,RHBrev}
for reviews), it is an important
endeavor to distinguish between generic cosmic strings and cosmic 
superstrings. As we shall demonstrate in this note, the plethora 
of small scale Cosmic Microwave
Background (CMB) experiments, such as the ground based 
Atacama Cosmology Telescope (ACT) \cite{ACT} and the 
South Pole Telescope (SPT) \cite{SPT} or the Planck
satellite experiment \cite{Planck}, will provide the observational 
means to make this distinction.  

Both generic cosmic strings and cosmic superstrings leave imprints in
the CMB.  The key signature - the ``Kaiser-Stebbins (KS)
effect'' \cite{Kaiser} - consists of line discontinuities
in the temperature map formed from a combination of
gravitational lensing and the Doppler effect: photons from the last 
scattering surface streaming by either
side of a moving cosmic string will be observed to have a temperature
which differs by a small amount proportional to the string tension.
The signature is manifest in position space - in Fourier space
the phase information which is crucial to obtain a line discontinuity
is lost. Therefore, to find evidence for the line discontinuities,
new data analysis techniques are required which work directly in
position space. One such technique is the Canny edge detection 
algorithm \cite{Canny1,Canny2}.
As shown in \cite{Amsel,Stewart,Danos:2008fq}, the
Canny algorithm can be used to search for the line discontinuities
produced by strings. In fact, it was shown that when applied
to data with angular resolution comparable to that which will
be obtained from the South Pole Telescope, the non-detection
of KS lines will allow an improvement in the upper bound on
the cosmic string tension by almost an order of magnitude
compared to existing bounds. 

One essential distinction between strings in simple gauge
theory models and cosmic superstrings is that the latter  
come in different varieties \footnote{For sufficiently complicated
gauge theory models, it is also possible to obtain
different types of strings \cite{Bucher}. Fat strings
arising in theories with light moduli fields also can give
rise to junctions \cite{Wells} - in this case due
to mergings of strings with low winding number into strings
of higher winding number. For simple gauge theory strings,
strings with higher winding number are unstable to the fragmentation
into strings of winding number 1, and hence these junctions to
not arise.}. There can be fundamental strings
(F-strings), one space-dimensional branes (D-strings), and
bound states of the above. These flavors of cosmic superstrings 
can form junctions, points where three strings meet in the shape of a Y 
\cite{Copeland:2003bj,Leblond:2004uc,Jackson:2004zg,Hanany:2005bc,
Hashimoto:2005hi,Copeland:2006eh,Copeland:2006if,Copeland:2005cy,Tye:2005fn}.

In the present paper we find that
the Canny algorithm can be modified in order to allow a
differentiation between string maps with and without string junctions.
Since string junctions are generically predicted in models
of brane inflation with cosmic superstrings but are not present
for cosmic strings in simple gauge theory models, our work
points towards a way of finding observational evidence which
would favor cosmic superstrings over simple gauge theory strings. 

To identify the line discontinuities predicted by cosmic
strings, good angular resolution of a CMB experiment is
crucial. It is for this reason that our simulations work
with maps similar in angular extent and angular resolution
which will be obtained by the ACT and SPT experiments, those
with ideal angular resolution. 

Our implementation of the Canny algorithm \cite{Danos:2008fq}
finds line discontinuities produced by the KS effect and produces
a map of edges. In contrast to previous work \cite{limits} on constraining
the cosmic string tension from CMB observations which focused on
the angular power spectrum of CMB maps in which the KS signature 
is washed out, our work searches specifically for the line 
discontinuities predicted by the KS effect. 

Our numerical work is based on simulations of the
predicted CMB anisotropies in small patches of the sky (side
length $10^{o}$) with angular resolution of $1.5^{'}$.
We simulate maps predicted in pure string models (i.e. no
``Gaussian noise'') with and without junctions, and also
maps in which the ``Gaussian noise'' due to inflationary
perturbations dominates the total amplitude of the power
spectrum, and cosmic strings contribute a sub-dominant
fraction consistent with the current upper bounds \cite{limits}.
The current upper bound on the contribution of strings to
the angular power spectrum of CMB anisotropies corresponds to
a value of the string tension $\mu$ which is $G \mu < 2 \times 10^{-7}$
in dimensionless units ($G$ is Newton's gravitational constant).

Our implementation of the Canny algorithm consists of a set of
routines which generate a map of edges in the sky corresponding 
to gradients in CMB maps which are in the range expected
from the KS effect \cite{Danos:2008fq}, and statistically
analyze the resulting edge maps, with the goal of
distinguishing between simulated CMB maps of pure Gaussian
inflationary perturbations and maps combining simulated cosmic strings
with Gaussian perturbations.  
In our studies \cite{Danos:2008fq} we showed that this distinction
could be made up to a three sigma confidence level for strings with 
tensions as low as $2\times 10^{-8}$ for maps without the removal of 
point source noise (unsmoothed) and
as low as $9\times 10^{-8}$ for maps with point source noise removed 
(smoothed).

Here we extend the use of the Canny algorithm to distinguish simulated maps
of strings with and without junctions. Based on the edge maps which
the Canny algorithm produces, we count the 
number of edges present in groups of pixels, thus determining the 
density of edges.  We then use the distribution of densities
to differentiate between maps with and without junctions.
Specifically, we compare the shapes of the histograms (number of occurrences
versus number of edges) in the two cases, using the ``combined Fisher
probability method''. 

The simulations contain various free parameters. Firstly, there is
the cosmic string tension $\mu$. Secondly, there is the number $N$
of string segments per Hubble volume per Hubble time during the
string scaling phase. The fraction of string segments which involve
a junction is another choice which must be made. In order to
ascertain that a deviation in the shape of an edge histogram from
that in a theory with pure Gaussian noise is a consequence of
strings with junctions, we must allow the parameters in the
simulations with and without junctions to be different. For example,
it is not sufficient to show that the histogram in a string simulation
with junctions for $N = 10$ is different from that of a string
simulation without junctions for the same value of $N$. We must
also check that the change in the shape resulting from the addition
of junctions cannot be masked by a change in the value of $N$.
We have carefully studied this point and find that, with appropriately
chosen number of boxes sampled in the observed window, we 
are able to differentiate string simulations with and without
junctions. Without Gaussian noise the difference in the shape of
the histogram is manifest. However, with Gaussian noise consistent
with the cosmic strings contributing not more than they are
allowed to by the limits of \cite{limits} we find that the
difference is no longer manifest. However, with sufficient care,
we are still able to make the statistical distinction.

The outline of this paper is as follows: we first describe our simulated
microwave sky patches resulting from models with cosmic strings junctions
(Section II).  In Section III we review the application of the Canny 
algorithm and present the edge maps which result from our simulations.  
We present the density distribution analysis in Section IV.
Finally, in Section V we conclude with a discussion of the results.

\section{Simulations}

We refer the reader to \cite{Danos:2008fq} for details on simulating the
combined maps produced by the combination of cosmic strings 
and inflationary perturbations. In brief, the contribution of
Gaussian inflationary perturbations to the CMB sky map is produced 
by Fourier transforming the angular power spectrum $C_l$ obtained
using the CAMB code \cite{CAMB}, assuming random phases.

The temperature map from pure cosmic strings is constructed using
an analytical toy model of the effects of strings on the microwave
sky. This model was first introduced in \cite{Periv} to model the
effects of long strings, and has been used widely since then. The
model assumes that the distribution of cosmic strings obeys a
scaling solution (i.e. the network looks the same at all times if
distances are scaled to the Hubble radius). The string network
consists of ``long'' strings (strings whose curvature radius is
comparable or larger than the Hubble radius) and loops (of radius
smaller than the Hubble length). Based on numerical simulations
of the evolution of a cosmic string network \cite{simuls} we
assume that the long strings dominate. We also assume that the
long strings are straight. Thus, according to the model of \cite{Periv},
the network of strings can be modelled as a set of straight string
segments of roughly Hubble length. The various cosmic string
simulations all agree that there is a scaling solution for the network
of long strings, but they do not agree on the density of strings.
We will model this uncertainty with a free parameter $N$, an integer
which gives the number of string segments per Hubble volume.
Since the strings move relativistically, the network of strings
looks uncorrelated on time scales larger than the Hubble time. Following
\cite{Periv}, we model this by taking the string segments to be
statistically independent in each Hubble time step.  
The strings are laid down in Hubble time steps from the time of 
recombination to the present time.

\begin{figure}
\includegraphics[scale=0.5]{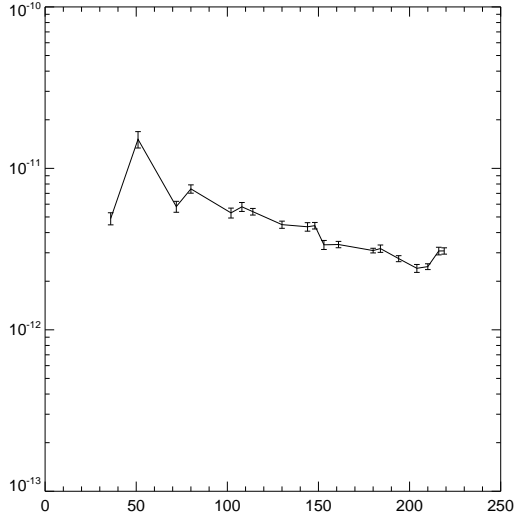}
\caption{The angular power spectrum of the CMB anisotropy maps of pure
cosmic string simulations with values N=10 and $G\mu=1\times 10^{-7}$.
The horizontal axis is $l$, the vertical axis is $l(l+1)C_l$.}
\label{clspectrumnojunc}
\end{figure}

Each string segment produces a characteristic rectangular temperature
fluctuation pattern in the sky which is obtained by
adding 
\be \label{basic}
\Delta T/T \, = \, 4\pi G\mu\tilde{v}r
\ee
on one side of the location of the string, affecting the sky to a
depth of a Hubble length \footnote{The range of the
``deficit angle'' \cite{deficit} in the plane perpendicular to the
string which leads to the KS effect is bounded from above by causality to
be the Hubble length - assuming that the cosmic strings are produced
in a phase transition in the early universe - and the work of
\cite{Joao} shows that the effect of the deficit angle in fact
extends just about to the causality limit.}
and subtracting this value from the other 
side. Here $\mu$ is the string tension, $\tilde{v} \equiv v\gamma(v)$ 
where $v$ is the transverse velocity of the string and its relativistic 
$\gamma$ factor is $\gamma(v)$, and $r$ is a random number between 0 
and 1 which is introduced 
to take into account projection effects and the fact that the
velocities of different strings will vary - in (\ref{basic}) we
use one fixed velocity $v$. We will speak in terms
of a pattern of rectangular ``boxes'' (one on each side of the string) 
which a string produces in the CMB sky. One 
difference between this simulation and that of 
\cite{Danos:2008fq} is that in the
former paper the length of the string segment laid down was the same length
as the depth of the box perpendicular to the string.  In this simulation we
allow the depth of the box affected by the string to be a different 
random fraction of the Hubble length (as determined by projection
effects).  
Our cosmic string toy model yields an approximately scale-invariant
spectrum of CMB anisotropies (as expected from early analytical work
on cosmic strings \cite{early}). The $C_l$ spectrum for 100 string
simulations of string tension $G\mu=1\times 10^{-7}$ is shown in 
Figure \ref{clspectrumnojunc}.

Since we have applications to small angular scale CMB experiments such
as ACT and SPT in mind, we take our window size to be 10 degrees and 
the resolution to be 1.5 arcminutes.

To sum the Gaussian and cosmic string maps we multiply the COBE-normalized
Gaussian map by a coefficient $a$ and add the cosmic string temperature map
obtained for the value of $G \mu$ which we are interested in:
\be
T_{(G+S)} \, = \, aT_{(G)} + T_{(S)}. 
\ee
The coefficient $a$ is obtained by demanding that the total map yields the
best fit to the observed angular CMB correlation function.

\begin{figure}
\includegraphics[scale=0.8]{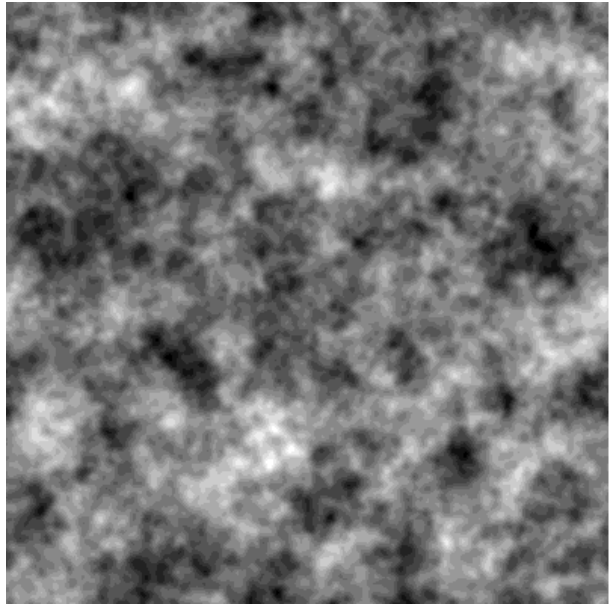}
\caption{Temperature map in a model with both cosmic strings ($N=10$ and
$G \mu = 10^{-7}$, no junctions) and Gaussian fluctuations.}
\label{N=10 strings + gaussian without junctions}
\end{figure}

We refer the reader to Figure \ref{N=10 strings + gaussian without junctions}
for the simulated map of cosmic strings (no junctions) plus Gaussian 
inflationary perturbations.
The map is for a value $G \mu = 10^{-7}$.

Next we describe the simulations in the presence of junctions. The
first difference is that not all strings have the same tension.
By $\mu_i$ we will denote the tension of the i'th string. The
vector ${\bf{\mu_i}}$ has magnitude $\mu_i$ and points in direction
of the string. We will
take the total number of string segments per Hubble volume to be
$N = 10$. We assume junctions in the shape of a ``Y'', i.e.
junctions with three legs each. Thus, there will be one string
segment per Hubble volume which is not part of a junction.

As in the
string simulation without junctions, the beginning of the first string in the
junction is placed randomly within a window extending by one Hubble length
in each direction around the observed window.  This accounts for strings 
which might begin within the field of view and extend out of it and for 
strings which might begin outside the field of view and extend into the
observed window.  The first string's starting point in this simulation will
be the point where the three strings meet.  The angle and sign of its 
velocity vector
as well as the direction of the string are randomly determined.  The string's
velocity angle and sign determine which of the two boxes around the string
will be assigned a positive and which a negative temperature fluctuation 
compared to the average. The formula (\ref{findtemperature}) taken 
from \cite{Brandenberger:2007ae} to determine
which side of the string has a positive temperature difference compared to the
average is:
\be \label{findtemperature}
\frac{\delta T}{T} \, = \, 8\pi G\gamma|\bm{v}\cdot(\bm{\mu_i}\times\bm{k})|
\ee
in which  $\bm{v}$ is the velocity vector of the string, and $\bm{k}$
is the line of sight unit vector perpendicular to the observation window. 

The first string segment of a junction has a fixed input string tension.  
We use $G\mu=8.7\times 10^{-8}$
because when there are three junctions per Hubble volume this gives a
coefficient $a=0.976$. Recall that $a$ is the coefficient 
used to sum the
Gaussian contribution of the map with the string map.  A value of
$a=0.976$ is obtained in the case of a simulation with $G\mu=1\times 10^{-7}$ 
for $N=10$ without junctions.
We want to compare different maps with the same contribution from the Gaussian
perturbations.  Therefore we fix
$a$ and compute the corresponding $G\mu$ for simulations with junctions
and varying values of $N$.  To determine the value of $G \mu$ for
a simulation with junctions, we first perform a simulation with a
given value $G\mu_0$ and compute the resulting value of $a$, denoted here
by $a_0$. The value of $G \mu$ which yields the required value of $a$
is then given by:
\be
G\mu^2 \, = \, \left(\frac{a^2-1}{a_0^2-1}\right)(G\mu_0^2) \, .
\ee
The value of $a$ for a given string tension and given value of $N$ will be 
higher than in our previous paper \cite{Danos:2008fq} since the boxes 
of altered temperature per the KS effect are smaller in this simulation 
where the depth can be less
than the string width.  Hence, the strings contribute less power.

\begin{figure}
\includegraphics[scale=0.8]{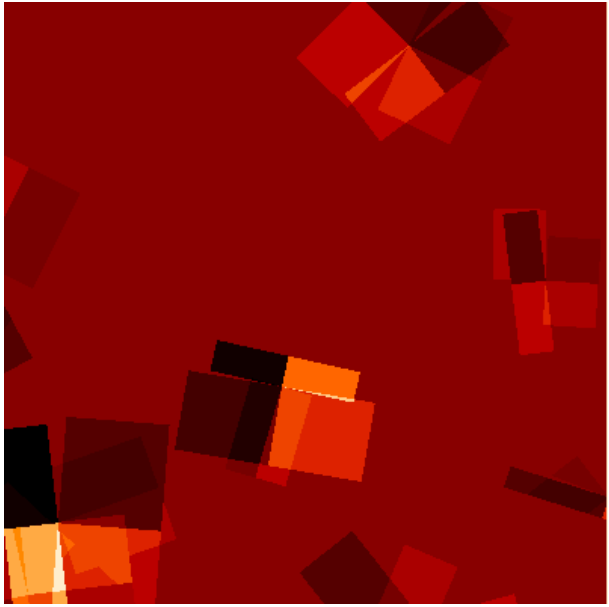}
\caption{A few cosmic strings with junctions.}
\label{teststringjuncmap}
\end{figure}

When simulating junctions, we draw the second string at the same point as the 
beginning of the first string with a string tension given by a random 
fraction of the first string's tension.  The direction of the second string
is random.  The third string's direction is chosen such that the vector sum
equals zero for 
the three vectors with magnitudes given by the string tensions and directions
given by the directions of the strings.  The magnitude of this third vector
indicating the direction of the third string gives the string tension for the
third string.  Figure \ref{teststringjuncmap} demonstrates the temperature
signature for a few strings with junctions.

\begin{figure}                                                                 
\includegraphics[scale=0.5]{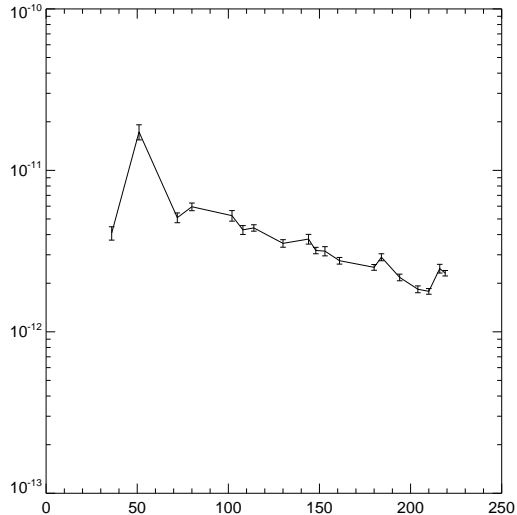}
 \caption{The angular power spectrum of the CMB anisotropy maps for          
cosmic string simulations with three junctions per Hubble volume and a
scaling solution with $N=10$ and $G\mu=8.7\times 10^{-8}$.       
The horizontal axis is $l$, the vertical axis is $l(l+1)C_l$.}                
\label{clspectrumjunc}                                                      
\end{figure}

As in our previous paper \cite{Danos:2008fq} we lay down
strings independently in each Hubble expansion time interval
starting from recombination until the present time. 
Figure \ref{clspectrumjunc} shows the resulting angular power
spectrum ($C_l$ spectrum) for
a scaling solution of $N=10$, a string tension of the first string as 
$G\mu=8.7\times 10^{-8}$ and three junctions per Hubble volume.
The value of the string tension without junctions
was chosen to lie slightly below the
current best upper limits; the string tension for simulations with junctions
was then selected to have the same contribution of the background Gaussian
perturbations.  

\section{Canny Algorithm}

\begin{figure}
\includegraphics[scale=0.8]{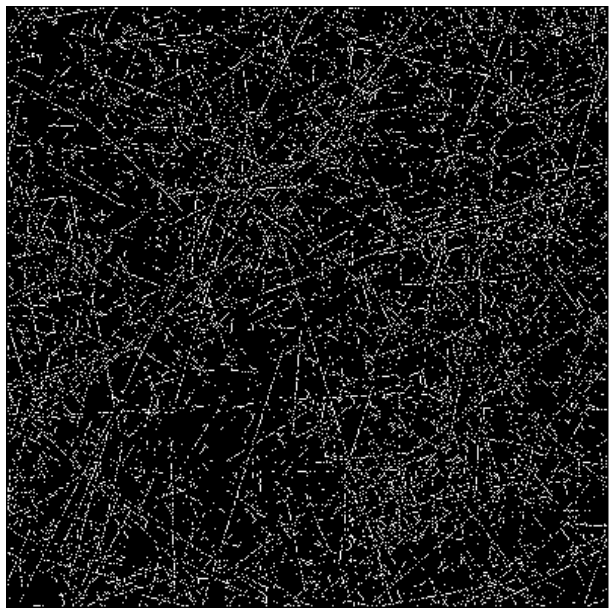}
\caption{Pure string edge map without junctions obeying a scaling solution
of N=10.}
\label{nojuncedgemap}
\end{figure}

\begin{figure}
\includegraphics[scale=0.8]{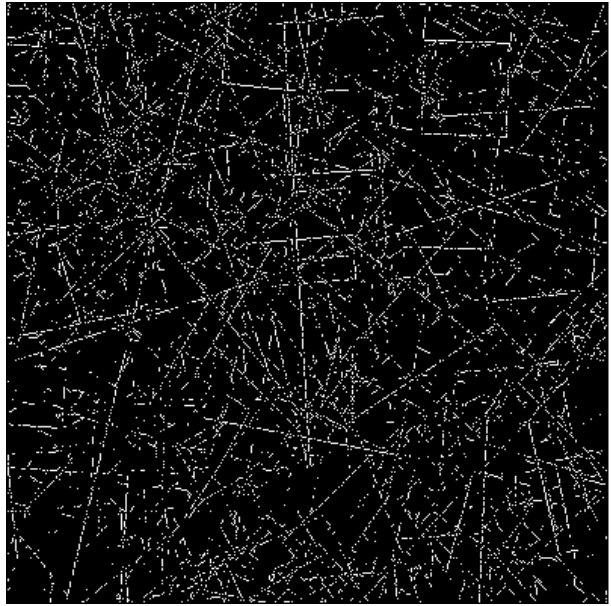}
\caption{Cosmic string edge map with three junctions per Hubble volume obeying
a scaling solution of N=10.}
\label{juncedgemap}
\end{figure}

We refer the reader to \cite{Danos:2008fq} for a thorough discussion of our
application of the Canny algorithm.  In essence, we look for edges by
searching for maximal gradients larger than a threshold $t_u$.  These
local maxima are perpendicular to the direction of the edge.  Gradients
larger than a cutoff value, $t_c$, are removed as these will likely be due
to the underlying Gaussian perturbations.  Gradients that are between the
upper threshold and a lower threshold are marked with a value of one
and counted as part of an edge
if they are connected to a point in a satisfactory direction (perpendicular
or near perpendicular)
whose gradient is a local maximum greater than or equal to the upper 
threshold.  Our algorithm creates a map of edges, or a map of points which
are marked as one connected together in edges. These can be seen for strings
without junctions obeying a scaling solution of $N=10$ in Figure
\ref{nojuncedgemap} and strings with three
junctions per Hubble volume obeying a scaling solution of $N=10$ in 
Figure \ref{juncedgemap}. Once an edge map has been created the program 
counts the number of continuous points that satisfy the above criteria, 
thus counting the number of points in the edges.

In our simulations, we have chosen the following values of the
cutoff parameters which enter the Canny algorithm: the top cutoff
is $t_c = 3$ \footnote{In units of the maximal gradient which can
be produced in the simulation by a cosmic string.}, 
the upper cutoff is $t_u =0.25$ and the lower cutoff is
$t_l =0.1$. These values were chosen based on the parameter optimization
performed in our previous work \cite{Danos:2008fq}.

\section{Density Distribution Analysis}

Because strings with junctions are grouped together, we hypothesized that 
the distribution of edges with junctions would differ from the distribution
without junctions.  We divided the observation window by a grid into boxes 
and counted the number of edges and number of points marked as one (the number
of pixels in edges) per grid box. The size of a grid box is a free 
parameter in our analysis.
We then plotted the number of boxes versus the number of either edges or 
points marked as one per box.  Using the number of edges
or the number of points marked as one per box gives slightly different 
statistics.  

When we count the number of edges in a 
given grid box, we need an algorithm to determine which box an edge falls
within when edges cross boxes.  We choose to assign an entire edge to the box
containing a specific pixel of that edge.
We could use the first pixel 
as the point denoting the edge, but this introduces a position bias because
the algorithm counts edges starting from the bottom left of the window.
However, when searching for pixels to include in an edge, the algorithm does
not search in a linear fashion so the final pixel included in an edge does not
have a biased position relative to the other pixels in the edge.  
Therefore, we assign the position of the edge to the final pixel.

We now discuss our numerical results. All of the simulations
were for the same value of $a$.
In all the following figures the blue
(solid) curve is for $N=10$ with three junctions per Hubble volume.
The other curves correspond to cosmic string simulations without junctions, 
the red innermost (dash dot dot) curve is for $N=1$,  
the yellow (dashed) one is for $N=6$, 
the green (dot-dashed) one is for $N=5$, 
and the outermost red (dotted) one is for $N=10$.
All curves correspond to the average over 100 edge maps. Note that the
$N = 10$ curve with junctions is closest to the $N = 5$ curve without
junctions.

The statistic we use to determine if the curves are distinguishable is a
combination of the Student t-test and the Fisher combined probability test.
For each value of the x-axis (the number of points marked as one or the number
of edges) we are given the mean number of boxes that contain that number of
points/edges and the standard deviation.  With two means and corresponding
standard deviations we can apply the t-test to compute the probability $p_l$
that the two means originate from the same distribution.  The Fisher combined
probability method then computes $\chi^2$ as follows:
\be
\chi^2_{2k} \, = \, -2\sum_{l=1}^k\ln(p_l)
\ee
where $k$ is the maximal number of edges/points marked as one being 
which arise.  We compute the probability value from the $\chi^2$ 
distribution with $2k$ values. Bins with entries or standard
deviations equal to zero are obviously not included in the sum.

\begin{figure}
\includegraphics[scale=0.5]{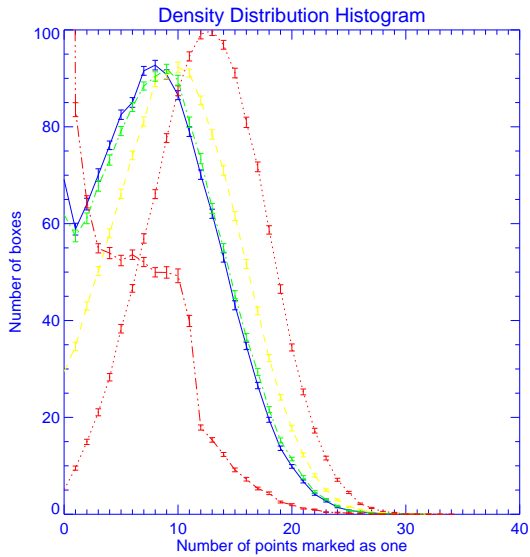}
\caption{Pure string histograms for 36$\times$36 boxes. Number of boxes
versus the number of points marked as one.  The red innermost curve (dash
dot dot) is the 
average over 
100 edge maps for $N=1$, the blue
curve (solid) 
is for $N=10$ with three junctions per Hubble volume, the yellow curve
(dashed) is
for $N=6$, the green curve (dash dot) is for $N=5$, and the 
outermost red curve (dotted) is for 
$N=10$.  Only the blue curve includes the presence of junctions.  See Table 1
for string tensions and $a$ coefficient.}
\label{densityhist.S.b36.b36}
\end{figure}

\begin{figure}
\includegraphics[scale=0.5]{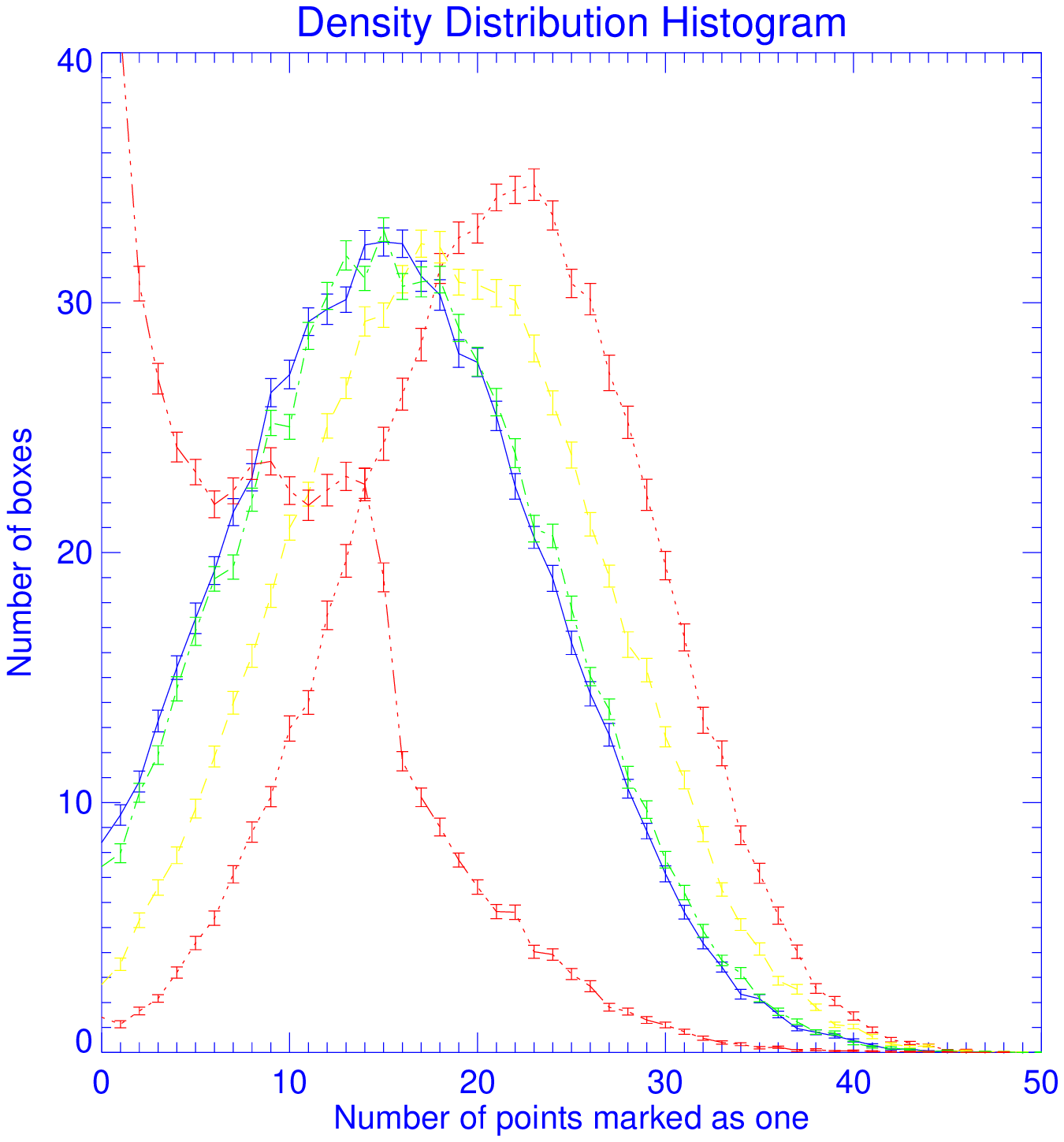}
\caption{Pure string histograms for 26$\times$26 boxes. Number of boxes
versus the number of points marked as one.  
The red innermost curve (dash
dot dot) is the 
average over 
100 edge maps for $N=1$, the blue
curve (solid) 
is for $N=10$ with three junctions per Hubble volume, the yellow curve
(dashed) is
for $N=6$, the green curve (dash dot) is for $N=5$, and the 
outermost red curve (dotted) is for 
$N=10$.  Only the blue curve includes the presence of junctions. }
\label{densityhist.S.b26.b26}
\end{figure}

\begin{figure}
\includegraphics[scale=0.5]{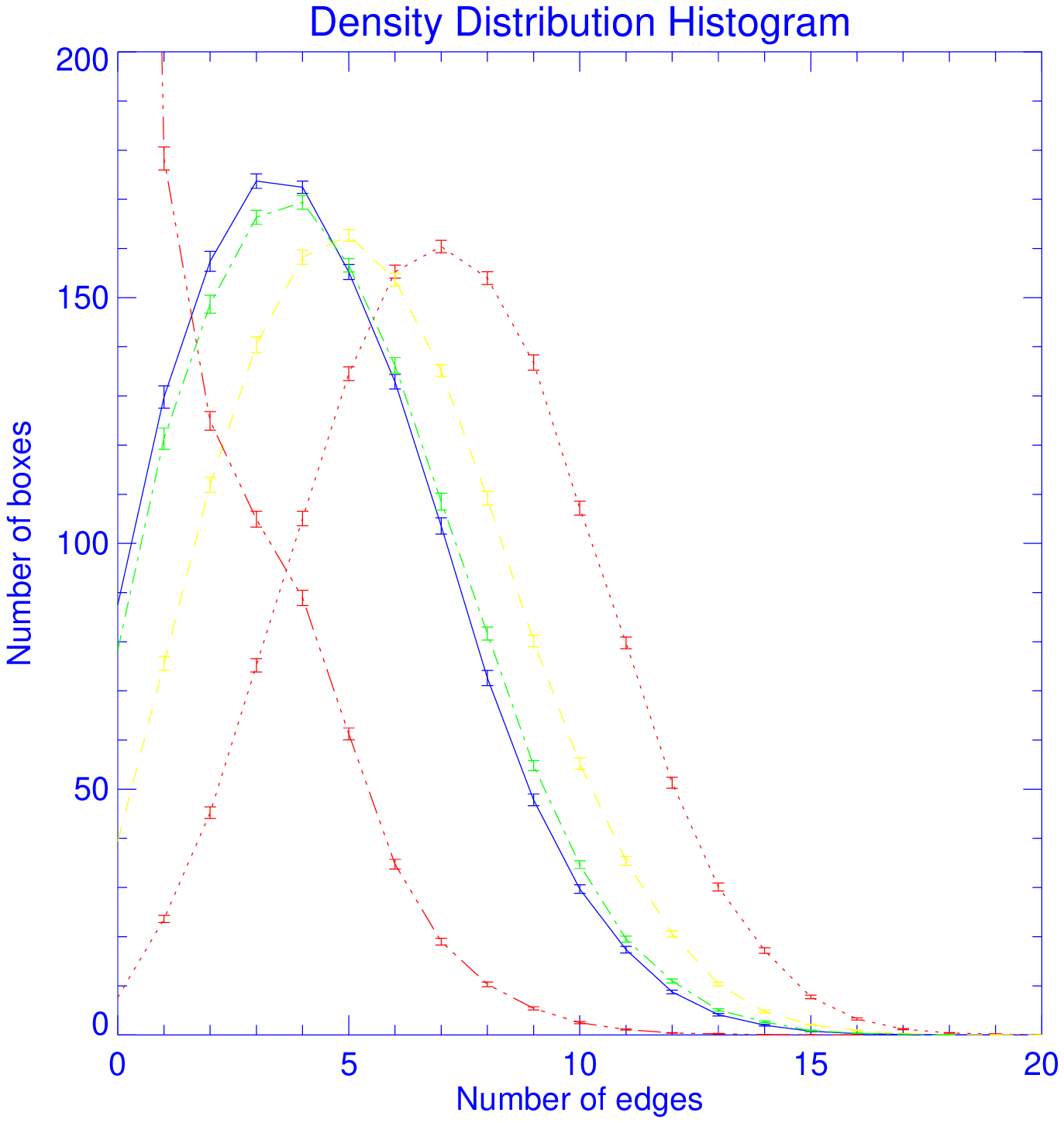}
\caption{Pure string histograms for 36$\times$36 boxes. Number of boxes
versus the number of edges.  
The red innermost curve (dash
dot dot) is the 
average over 
100 edge maps for $N=1$, the blue
curve (solid) 
is for $N=10$ with three junctions per Hubble volume, the yellow curve
(dashed) is
for $N=6$, the green curve (dash dot) is for $N=5$, and the 
outermost red curve (dotted) is for 
$N=10$.  Only the blue curve includes the presence of junctions. }
\label{edgedensityhist.S.b36.b36}
\end{figure}

\begin{figure}
\includegraphics[scale=0.5]{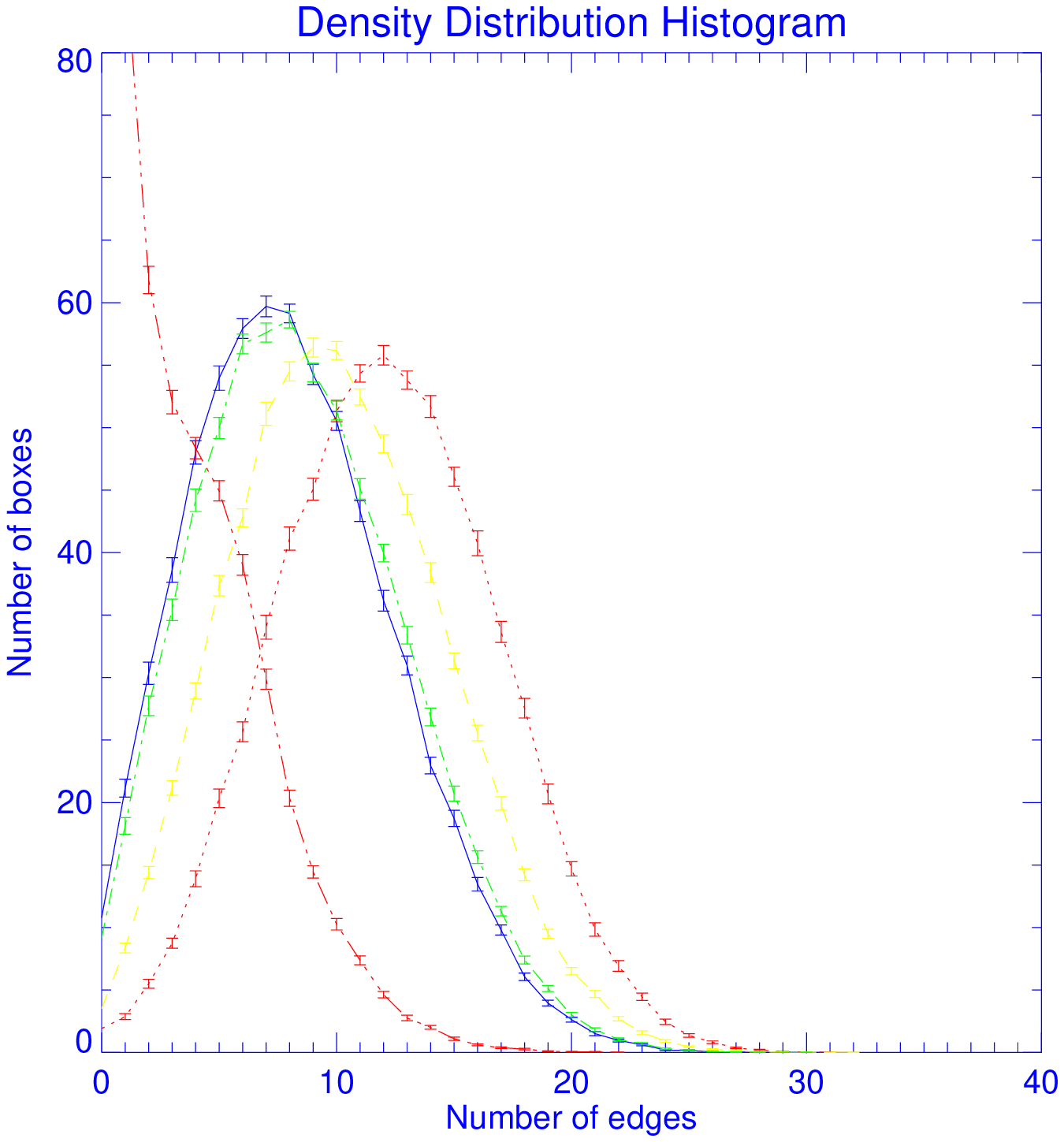}
\caption{Pure string histograms for 26$\times$26 boxes. Number of boxes
versus the number of edges.  
The red innermost curve (dash
dot dot) is the 
average over 
100 edge maps for $N=1$, the blue
curve (solid) 
is for $N=10$ with three junctions per Hubble volume, the yellow curve
(dashed) is
for $N=6$, the green curve (dash dot) is for $N=5$, and the 
outermost red curve (dotted) is for 
$N=10$.  Only the blue curve includes the presence of junctions. }
\label{edgedensityhist.S.b26.b26}
\end{figure}

Figures \ref{densityhist.S.b36.b36}-\ref{edgedensityhist.S.b26.b26} are 
pure string (no Gaussian added) histograms.  We see that for the different
scaling solutions the peak and the width of the curves are shifted.  We also 
see that it is difficult to distinguish by eye between the scaling 
solution $N=5$ without junctions and $N=10$ with junctions, but that 
using the Fisher probability method these curves are clearly distinguishable.  

We also see that as we
change the box size (or the number of boxes in the grid) the results differ.
This opens up the possibility to probe the difference between two sky
maps using a set of different statistics. One statistic may be more
powerful at distinguishing the map with string junctions from a map
without junctions for one value of $N$, while another statistic may
yield a stronger discrimination for another value of $N$.

In Figure \ref{densityhist.S.b36.b36} we divide the 
observation window into 36 by 36 size boxes and plot the number of boxes
for pure string maps (without
Gaussian contribution) versus the number of {\it points marked as one} 
in each box. Table 1 shows the corresponding probabilities for the junction 
maps to have come from the same distribution as the corresponding string maps 
without junctions for various scaling solutions. In the table, each
line corresponds to a different comparison. The first column gives
the parameters of the simulation with junctions (which are always
taken to be the same), the second column states the parameters of
the simulation without junctions which the run indicated in the
first column is compared to. The third line gives the probability
that the two maps come from the same distribution. A probability of zero
means that the numerical result is smaller than the numerical cutoff. 

{F}igure \ref{densityhist.S.b26.b26} plots the same as above except that 
the grid is divided into 26 by 26 boxes.  Table 2 displays the results for 
this figure.

{F}igure \ref{edgedensityhist.S.b36.b36} plots the histogram of the number of 
boxes versus the number of {\it edges} for 36 by 36 boxes in the grid.  Table 3
displays the statistical results for this figure. We see that using
{\it edges} instead of {\it number of points} yields an improved
discriminatory power.
  
{F}igure \ref{edgedensityhist.S.b26.b26} plots the histogram for the number of
boxes versus the number of edges for 26 by 26 boxes in the grid with Table 4
giving the corresponding statistical results.

\begin{table}[h]
Table 1: Probabilities that Maps with Junctions come from same Distribution
as Maps without Junctions for Pure Strings\\
Based on the distribution of points marked as 1\\
$a=0.976$ 36x36 Boxes Per window\\
\begin{tabular}{|ccc|ccc|c|}
\hline
$N$&$G\mu$&junctions&$N$&$G\mu$ & junctions & probability\\
\hline
10&$8.7\times 10^{-8}$&yes&10&$1\times 10^{-7}$&no&0\\
10&$8.7\times 10^{-8}$&yes&6&$1.2\times 10^{-7}$&no&0\\
10&$8.7\times 10^{-8}$&yes&5&$1.37\times 10^{-7}$&no&$5\times 10^{-9}$\\
10&$8.7\times 10^{-8}$&yes&1&$3.12\times 10^{-7}$&no&0\\
\hline
\end{tabular}
\end{table}

\begin{table}[h]
Table 2: Probabilities that Maps with Junctions come from same Distribution
as Maps without Junctions for Pure Strings\\
(Based on the distribution of points marked as 1)\\
$a=0.976$ 26x26 Boxes Per window\\
\begin{tabular}{|ccc|ccc|c|}
\hline
$N$&$G\mu$&junctions&$N$&$G\mu$ & junctions & probability\\
\hline
10&$8.7\times 10^{-8}$&yes&10&$1\times 10^{-7}$&no&0\\
10&$8.7\times 10^{-8}$&yes&6&$1.2\times 10^{-7}$&no&0\\
10&$8.7\times 10^{-8}$&yes&5&$1.37\times 10^{-7}$&no&$3\times 10^{-6}$\\
10&$8.7\times 10^{-8}$&yes&1&$3.12\times 10^{-7}$&no&0\\
\hline
\end{tabular}
\end{table}

\begin{table}[h]
Table 3: Probabilities that Maps with Junctions come from same Distribution
as Maps without Junctions for Pure Strings\\
(Based on the distribution of edges)\\
$a=0.976$ 36x36 Boxes Per window\\
\begin{tabular}{|ccc|ccc|c|}
\hline
$N$&$G\mu$&junctions&$N$&$G\mu$ & junctions & probability\\
\hline
10&$8.7\times 10^{-8}$&yes&10&$1\times 10^{-7}$&no&0\\
10&$8.7\times 10^{-8}$&yes&6&$1.2\times 10^{-7}$&no&0\\
10&$8.7\times 10^{-8}$&yes&5&$1.37\times 10^{-7}$&no&0\\
10&$8.7\times 10^{-8}$&yes&1&$3.12\times 10^{-7}$&no&0\\
\hline
\end{tabular}
\end{table}

\begin{table}[h]
Table 4: Probabilities that Maps with Junctions come from same Distribution
as Maps without Junctions for Pure Strings\\
(Based on the distribution of edges)\\
$a=0.976$ 26x26 Boxes Per window\\
\begin{tabular}{|ccc|ccc|c|}
\hline
$N$&$G\mu$&junctions&$N$&$G\mu$ & junctions & probability\\
\hline
10&$8.7\times 10^{-8}$&yes&10&$1\times 10^{-7}$&no&0\\
10&$8.7\times 10^{-8}$&yes&6&$1.2\times 10^{-7}$&no&0\\
10&$8.7\times 10^{-8}$&yes&5&$1.37\times 10^{-7}$&no&$8\times 10^{-16}$\\
10&$8.7\times 10^{-8}$&yes&1&$3.12\times 10^{-7}$&no&0\\
\hline
\end{tabular}
\end{table}

\begin{figure}
\includegraphics[scale=0.5]{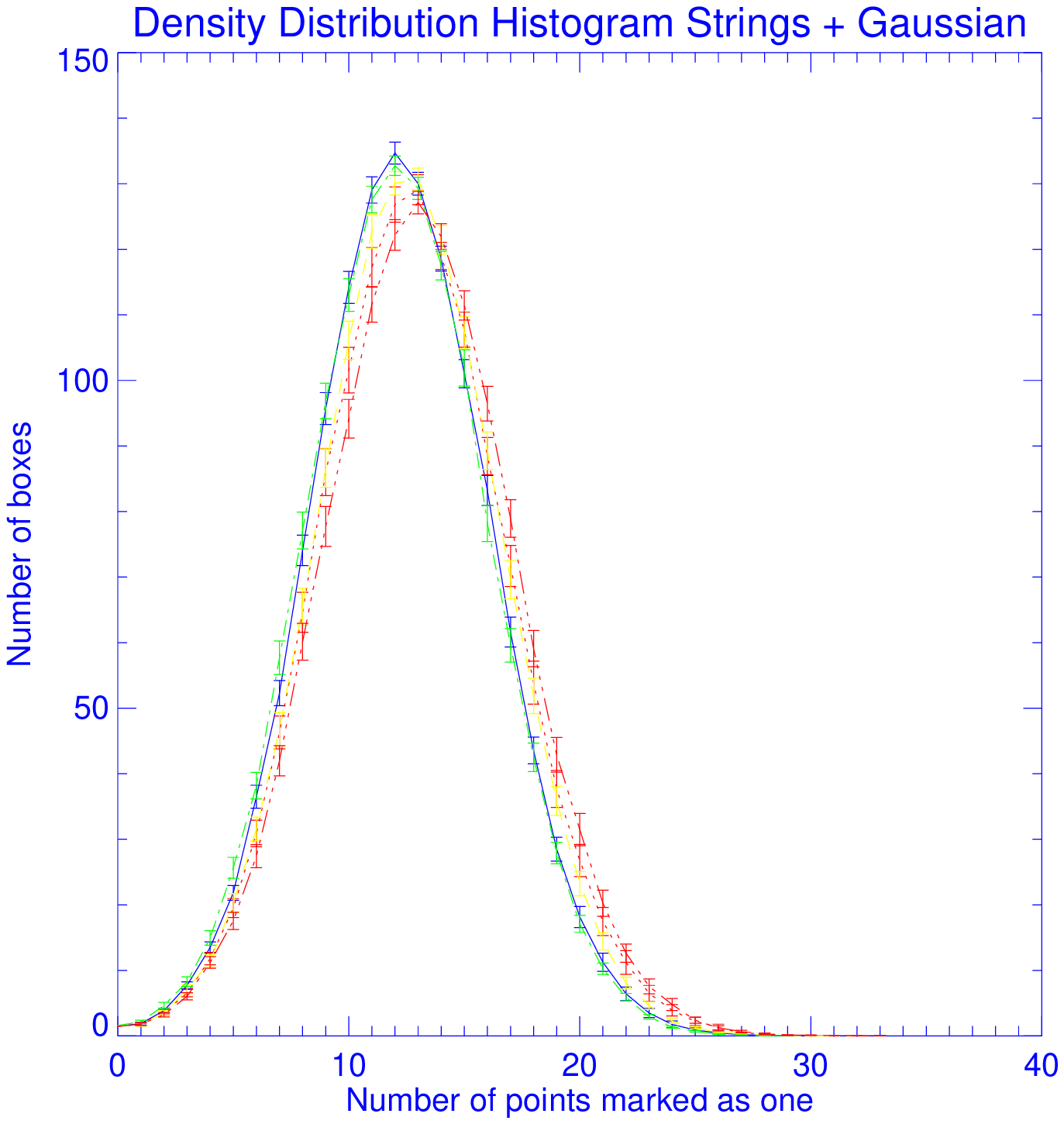}
\caption{Strings plus Gaussian histogram for 36$\times$36 boxes. Number of 
boxes versus the number of points marked as one.  
The red curve (dash
dot dot) is the 
average over 
100 edge maps for $N=1$, the blue
curve (solid) 
is for $N=10$ with three junctions per Hubble volume, the yellow curve
(dashed) is
for $N=6$, the green curve (dash dot) is for $N=5$, and the 
outermost red curve (dotted) is for 
$N=10$.  Only the blue curve includes the presence of junctions.}
\label{densityhist.SG.b36.b36}
\end{figure}

\begin{figure}
\includegraphics[scale=0.5]{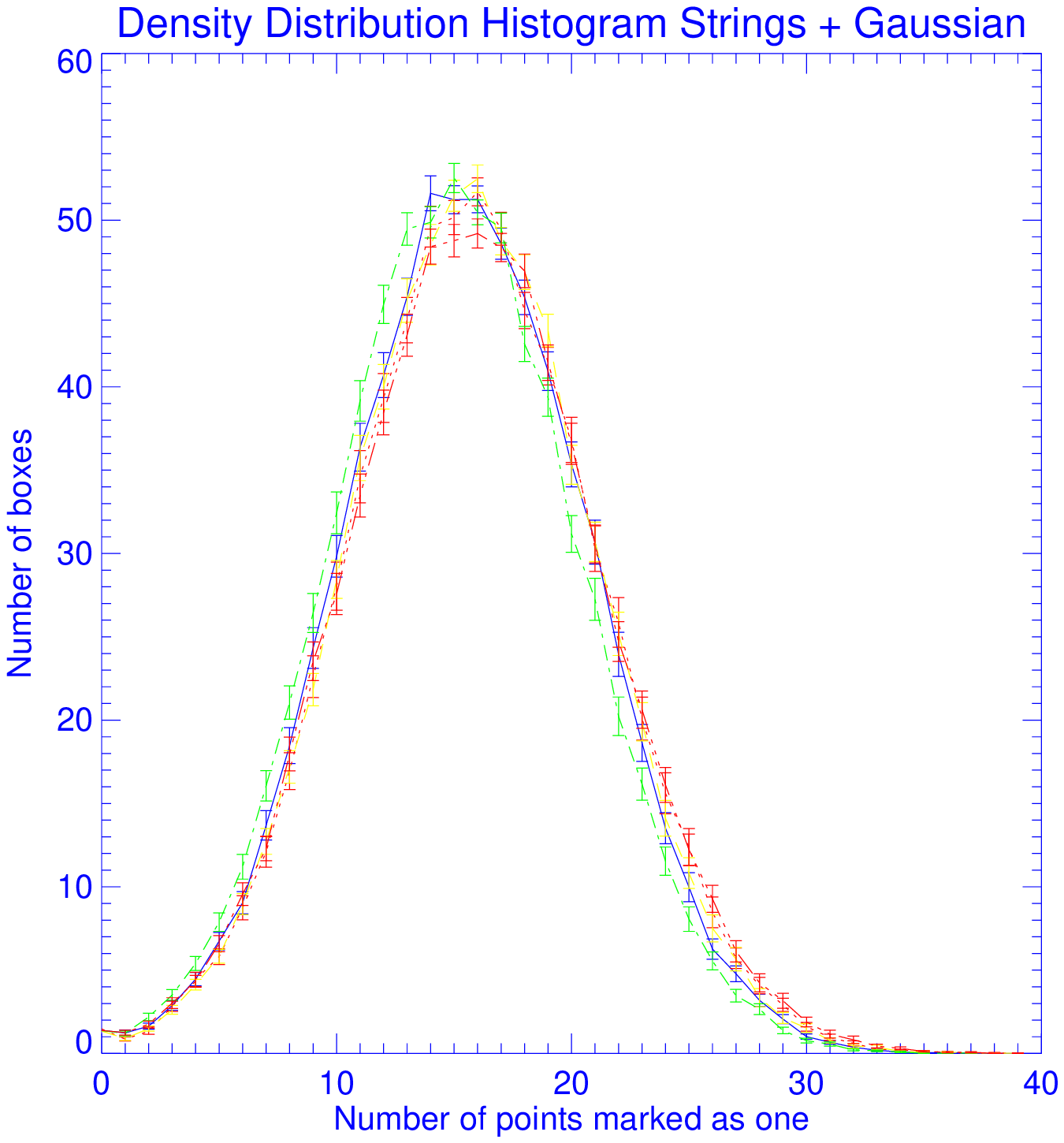}
\caption{String plus Gaussian histogram for 26$\times$26 boxes. Number of boxes
versus the number of points marked as one.  
The red curve (dash
dot dot) is the 
average over 
100 edge maps for $N=1$, the blue
curve (solid) 
is for $N=10$ with three junctions per Hubble volume, the yellow curve
(dashed) is
for $N=6$, the green curve (dash dot) is for $N=5$, and the 
outermost red curve (dotted) is for 
$N=10$.  Only the blue curve includes the presence of junctions. }
\label{densityhist.SG.b26.b26}
\end{figure}

\begin{figure}
\includegraphics[scale=0.5]{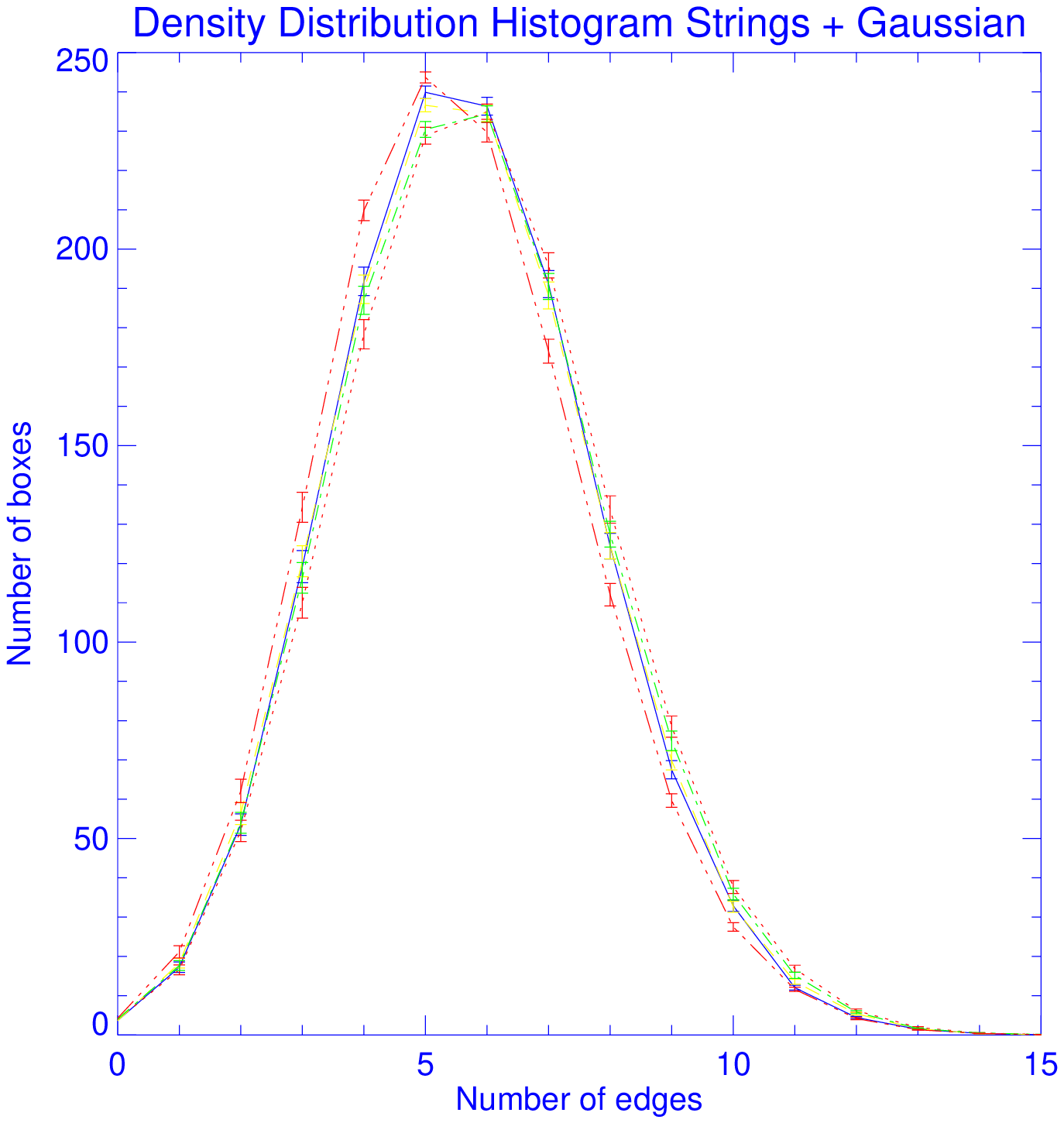}
\caption{String plus Gaussian histogram for 36$\times$36 boxes. Number of boxes
versus the number of edges.  
The red curve (dash
dot dot) is the 
average over 
100 edge maps for $N=1$, the blue
curve (solid) 
is for $N=10$ with three junctions per Hubble volume, the yellow curve
(dashed) is
for $N=6$, the green curve (dash dot) is for $N=5$, and the 
outermost red curve (dotted) is for 
$N=10$.  Only the blue curve includes the presence of junctions. }
\label{edgedensityhist.SG.b36.b36}
\end{figure}

\begin{figure}
\includegraphics[scale=0.5]{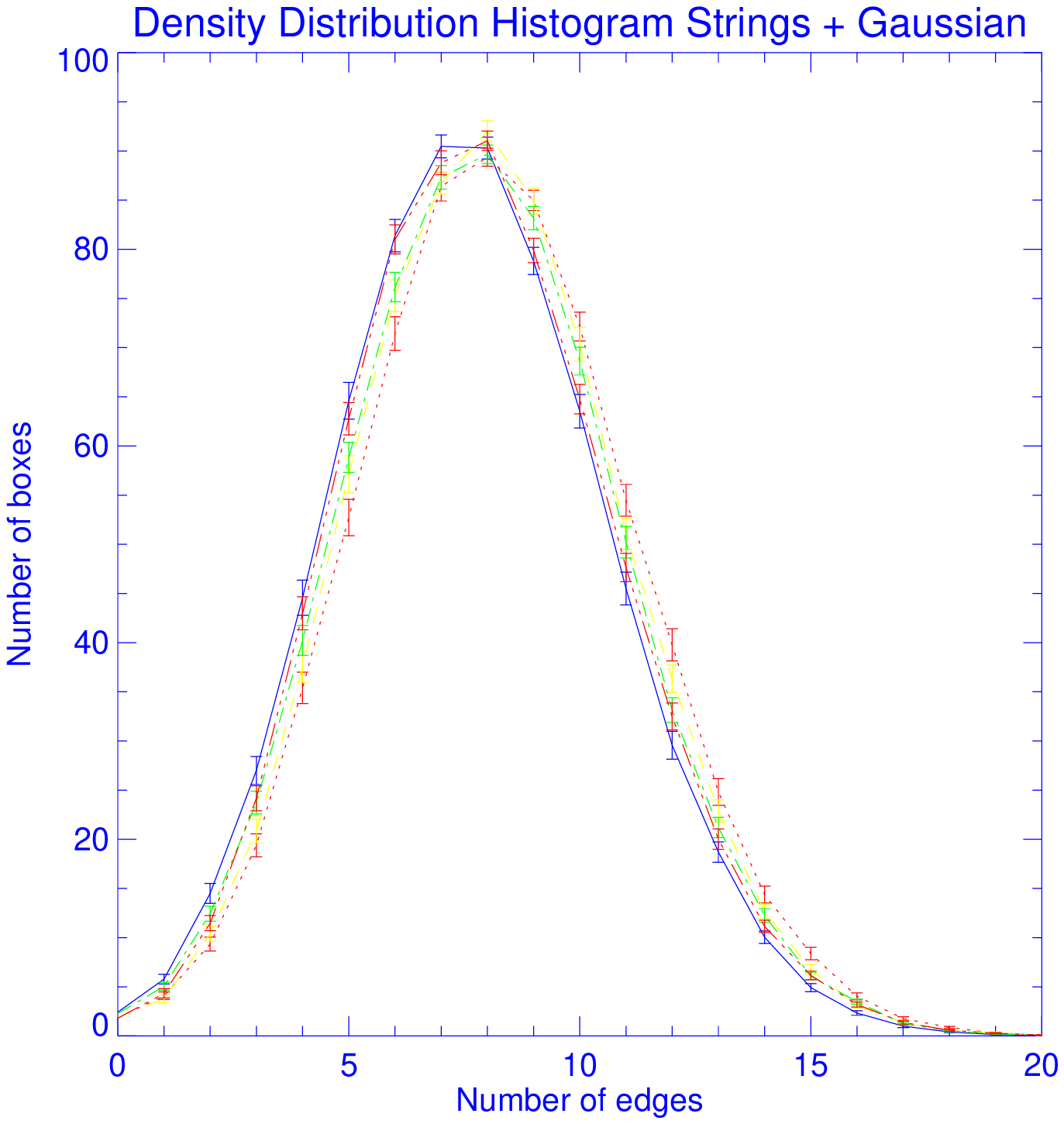}
\caption{String plus Gaussian histogram for 26$\times$26 boxes. Number of boxes
versus the number of edges. The red curve (dash
dot dot) is the 
average over 
100 edge maps for $N=1$, the blue
curve (solid) 
is for $N=10$ with three junctions per Hubble volume, the yellow curve
(dashed) is
for $N=6$, the green curve (dash dot) is for $N=5$, and the 
outermost red curve (dotted) is for 
$N=10$.  Only the blue curve includes the presence of junctions. }
\label{edgedensityhist.SG.b26.b26}
\end{figure}

{F}igures \ref{densityhist.SG.b36.b36}-\ref{edgedensityhist.SG.b26.b26} 
with corresponding Tables 5-8 give the results for the string maps with
Gaussian added.  The string tensions are adjusted so that the $a$ coefficient
remains the same for each scaling solution, hence keeping the percentage of
the Gaussian contribution constant.  We choose the baseline $a$ coefficient 
to be consistent with the string tension limits given by Pogosian, Wasserman, 
and Wyman (see the corresponding reference in the list \cite{limits}).

In Figure \ref{densityhist.SG.b36.b36}
the number of boxes versus number of points marked as one is plotted for
36 by 36 size boxes in a grid.  We can easily distinguish between scaling 
solutions of $N=10$, $N=6$, and $N=1$ of maps without junctions compared 
to maps with three junctions per Hubble volume.  With fixed 36 by 36 
boxes in a 10 degree window, maps with $N=5$ are indistinguishable
from maps of $N=10$ with three junctions per Hubble volume, as can be seen in
Table 5.  

However, here is where our power of being able to use a set of different
statistics comes in handy.
In Figure \ref{densityhist.SG.b26.b26}, corresponding to Table 6, we
use 26 by 26 size
boxes in a grid. Now, the results for $N=5$ are distinguishable from 
those in simulations with junctions, but the 
$N=10$ junction simulation results and the $N=6$ no junction scaling 
solution are no longer distinguishable.
With the combined results for both 36 by 36 and 26 by 26 size boxes 
we can distinguish maps with junctions from maps without junctions for 
strings plus Gaussian fluctuations for all values of $N$. The
explanation for our results is that the edges cluster
differently on different length scales for simulations with and without
junctions.  

A slightly different analysis is to consider the distribution of edges versus
the number of pixels in edges.  
With junctions we expect the edges to be longer and therefore less dense
in number than
the number of edges for maps without junctions.
Figure \ref{edgedensityhist.SG.b36.b36} and its
corresponding Table 7 indicate the results for number of boxes versus number
of edges for 36 by 36 size boxes.  From these results maps with junctions are
indistinguishable from maps of a no-junction scaling solutions with $N=6$. 
However, Figure
\ref{edgedensityhist.SG.b26.b26} and its corresponding Table 8 show that by
plotting the number of boxes versus number of edges for 26 by 26 size boxes, 
maps with junctions become distinguishable from maps without junctions 
for all scaling solutions we considered.

\begin{table}[h]
Table 5: Probabilities that Maps with Junctions come from same Distribution
as Maps without Junctions for Strings with Gaussian\\
(Based on the distribution of points marked as 1)\\
$a=0.976$ 36x36 Boxes Per window\\
\begin{tabular}{|ccc|ccc|c|}
\hline
$N$&$G\mu$&junctions&$N$&$G\mu$ & junctions & probability\\
\hline
10&$8.7\times 10^{-8}$&yes&10&$1\times 10^{-7}$&no&$3.2\times 10^{-13}$\\
10&$8.7\times 10^{-8}$&yes&6&$1.2\times 10^{-7}$&no&$3.6\times 10^{-6}$\\
10&$8.7\times 10^{-8}$&yes&5&$1.37\times 10^{-7}$&no&$0.79$\\
10&$8.7\times 10^{-8}$&yes&1&$3.12\times 10^{-7}$&no&0\\
\hline
\end{tabular}
\end{table}

\begin{table}[h]
Table 6: Probabilities that Maps with Junctions come from same Distribution
as Maps without Junctions for Strings with Gaussian\\
(Based on the distribution of points marked as 1)\\
$a=0.976$ 26x26 Boxes Per window\\
\begin{tabular}{|ccc|ccc|c|}
\hline
$N$&$G\mu$&junctions&$N$&$G\mu$ & junctions & probability\\
\hline
10&$8.7\times 10^{-8}$&yes&10&$1\times 10^{-7}$&no&0.069915751\\
10&$8.7\times 10^{-8}$&yes&6&$1.2\times 10^{-7}$&no&0.70260790\\
10&$8.7\times 10^{-8}$&yes&5&$1.37\times 10^{-7}$&no&$7.7\times 10^{-7}$\\
10&$8.7\times 10^{-8}$&yes&1&$3.12\times 10^{-7}$&no&0.00011939824\\
\hline
\end{tabular}
\end{table}

\begin{table}[h]
Table 7: Probabilities that Maps with Junctions come from same Distribution
as Maps without Junctions Strings with Gaussians\\
(Based on the distribution of edges)\\
$a=0.976$ 36x36 Boxes Per window\\
\begin{tabular}{|ccc|ccc|c|}
\hline
$N$&$G\mu$&junctions&$N$&$G\mu$ & junctions & probability\\
\hline
10&$8.7\times 10^{-8}$&yes&10&$1\times 10^{-7}$&no&$7.46\times 10^{-11}$\\
10&$8.7\times 10^{-8}$&yes&6&$1.2\times 10^{-7}$&no&0.76527496\\
10&$8.7\times 10^{-8}$&yes&5&$1.37\times 10^{-7}$&no&0.00028340182\\
10&$8.7\times 10^{-8}$&yes&1&$3.12\times 10^{-7}$&no&$4.2\times 10^{-10}$\\
\hline
\end{tabular}
\end{table}

\begin{table}[h]
Table 8: Probabilities that Maps with Junctions come from same Distribution
as Maps without Junctions for Strings with Gaussian\\
(Based on the distribution of edges)\\
$a=0.976$ 26x26 Boxes Per window\\
\begin{tabular}{|ccc|ccc|c|}
\hline
$N$&$G\mu$&junctions&$N$&$G\mu$ & junctions & probability\\
\hline
10&$8.7\times 10^{-8}$&yes&10&$1\times 10^{-7}$&no&0\\
10&$8.7\times 10^{-8}$&yes&6&$1.2\times 10^{-7}$&no&0\\
10&$8.7\times 10^{-8}$&yes&5&$1.37\times 10^{-7}$&no&$4.1\times 10^{-8}$\\
10&$8.7\times 10^{-8}$&yes&1&$3.12\times 10^{-7}$&no&0.00076553716\\
\hline
\end{tabular}
\end{table}

\section{Conclusions}

We have applied the Canny algorithm to simulated maps of CMB anisotropies
induced by models with cosmic stings either containing or not
containing string junctions. We proposed a statistic with which 
it should be possible to distinguish between the maps with and 
without junctions. This provides a method with which one can distinguish 
between the predictions of simple gauge field theory models with cosmic 
strings (which typically do not admit string junctions), and
models such as those giving rise to cosmic superstrings, which are
characterized by the presence of junctions.

We first showed that our statistic is able to clearly differentiate
between string maps with and without junctions in the absence of
Gaussian noise. However, since we know that a cosmic string model
without a dominant Gaussian contribution to the spectrum of
fluctuations is not consistent with the latest data on the angular
power spectrum of CMB anisotropies, we need to consider correctly
normalized sky maps which contain strings with a tension sufficiently
low such that its contribution to the power spectrum does not exceed
the current limits \cite{limits}, with the bulk of the power coming
from Gaussian noise. In this case it is more difficult to differentiate
between maps where the strings have junction and where they do not.
 
Nevertheless, making use of our detailed statistics, we found a clear 
distinction in the shape (peak and width)
of the density distribution (number of boxes sampled with varying numbers of
edges or pixels in the edges) for the different scaling solutions.  For 
example, $N=1$ strings per Hubble volume had a large number of boxes with just
a few edges, as would be expected, whereas $N=10$ strings per Hubble volume 
looked more Gaussian, as expected.  We found a 
statistical difference between the curves for junctions and the curves of
different scaling solutions without junctions.

Future work includes optimizing the Canny algorithm, and searching for
the lowest value of the string tension for which maps with and
without string junctions can be differentiated. It would also
be very interesting to algorithms to ``real'' string simulations
(those obtained via a numerical evolution of the Nambu-Goto
equations) rather than just relying on the simple toy models for
the distribution of strings which we have used. Maps of
CMB anisotropies produced by strings without junctions coming from
``real'' simulations are available \cite{Fraisse}. However,
to our knowledge there are no corresponding results for networks
with string junctions.  

\begin{acknowledgements}

This work is supported in part by an NSERC Discovery Grant and
by funds from the Canada Research Chairs Program. We would like
to thank David Morrissey and Andrew Frey for useful discussions.

\end{acknowledgements}


\end{document}